# The characteristics of LK-99 by $Cu_2S$ removal using ammonia solution : A diamagnetic semiconductor

**Zhujialei Lei[1,†], Chin-Wei Lin[1,†], I-Nan Chen[2,†], Chun-Tse Chou[1] & Li-Min Wang[1] ✉**

[1]Department of Physics/Graduate Institute of Applied Physics, National Taiwan University, Taipei 106, Taiwan

[2]Instrumentation Center, National Taiwan University, Taipei 106, Taiwan

[†] These authors contributed equally to this work.

✉email: liminwang@ntu.edu.tw.

**Abstract**

In this study, we re-evaluated the superconducting properties of LK-99. The LK-99 samples were synthesized using the process proposed by the original Korean team. Additionally, we examined whether the results of the Korean team are related to $Cu_2S$ by using ammonia solution ($NH_3·H_2O$) to remove $Cu_2S$. Through x-ray diffraction (XRD) analysis, a distinct $Cu_2S$ phase was identified in the LK-99 samples. A subsequent treatment using an ammonia solution effectively eliminated this phase. The appearance of blue $Cu^{+2}$ ions in the solution and the elimination of the $Cu_2S$ peak in XRD support the conclusion. The magnetic and electrical properties of LK-99 with and without $Cu_2S$ postulate that the superconducting-like behavior in LK-99 predominantly arises from a transition in resistivity due to the influence of $Cu_2S$. As such, LK-99 is better classified as a diamagnetic semiconductor than a room-temperature superconductor. The room-temperature superconductors still require further research.

**Introduction**

Realizing a room-temperature superconductor has always been a dream of humanity. Recently, a Korean research team announced the discovery of a room-temperature ambient-pressure superconductor: LK-99 [1,2], which caused excitement and scientific sensation worldwide. Theoretical calculations have also shown that under proper doping and lattice arrangement, LK-99 may exhibit superconductivity [3-6]. Sinéad M. Griffin, a researcher at the Lawrence Berkeley National Laboratory in the United States, pointed out that through computational simulations, the path of copper atoms infiltrating into the lattice is in the appropriate conditions and positions, *i. e.*, Cu substitution on the appropriate Pb(1) site, may let LK-99 display many vital characteristics for high-$T_c$ superconductivity, which is the first paper to prove the feasibility of the "LK-99" by theory[3]. Subsequently, other calculations of electronic band structures for LK-99 revealed by density functional theory (DFT) were also proposed [6-11]. DFT calculation results have shown that $Pb_9Cu(PO_4)_6O$ (the chemical formula of LK-99) has a flat energy band passing through the Fermi level, which in turn produces a high density of state (DOS) near the Fermi level, being one of the fundamental features known to give rise to superconductivity in the band-structure configuration. Therefore, the authors generally believe that the doping of Cu causes the transition from "insulator to conductor," and boldly infer that LK-99 may have superconducting properties. However, many research teams from various countries, mainly based on experimental work, tried to reproduce the results of the Korean research team's LK-99, but failed to prove that it is a room-temperature superconductor, but a semiconductor with diamagnetism [12-15].

One notable work by Zhu *et al*. argued that the so-called superconducting behavior in LK-99 is most likely due to a reduction in resistivity caused by the first-order structural phase transition of $Cu_2S$ at around 385 K, which could explain some of the unusual electrical properties of LK-99 [16]. Another work by Jain pointed out that the synthesized LK-99 may have a significant fraction of cuprous sulfide $Cu_2S$, which has a known phase transition at 104 ˚C from an ordered low-temperature phase to a high-temperature superionic phase, and thus exhibits sharp transitions in electrical resistivity and heat capacity [14]. This result coincides with the temperature-induced transitions reported for LK-99, and implies that LK-99 must be synthesized without any $Cu_2S$ to allow unambiguously confirming the superconducting properties of LK-99. Thus, a method of post-growth annealing on LK-99 was conducted to probe the impact of $Cu_2S$ [17]. In this research work, according to the disclosed synthesis method of LK-99, we tried for the first time to remove $Cu_2S$ phase in LK-99 by ammonia solution. With the x-ray diffraction (XRD) result, we found a significant removal of $Cu_2S$ phase in the sample via steeping it in an ammonia solution.

Furthermore, we were able to reproduce similar results of magnetic and electrical measurements demonstrated by the Korean team and found that the results needed to be corrected by considering the impact of $Cu_2S$. We conclude that $Pb_9Cu(PO_4)_6O$ is a diamagnetic semiconductor material, not a so-called room-temperature superconductor.

**Methods**

**Sample growth.** The synthesis process of $Pb_{10-x}Cu_x(PO_4)_6O$, LK-99, adheres to the method proposed by the Korean team [1,2]. First, the $Pb_2(SO_4)O$ is obtained by mixing PbO and $Pb(SO_4)$ in a 1:1 ratio and subjecting the mixture to a 725°C heat treatment for a full day in an atmosphere. In parallel, the Cu and P powders were amalgamating. This mixture was then heated to 550°C over for two days under a vacuum of $10^{-3}$ torr to produce $Cu_3P$. The resulting $Pb_2(SO_4)O$ and $Cu_3P$ were granulated and mixed. Then, they were enclosed within a vacuum-sealed quartz tube. Subsequently, the quartz tube underwent a heating process, reaching 925°C at a rate of 10°C/min and maintaining this temperature for 20 hours. After the heating period, it was cooled back to ambient conditions. Two different cooling rates were applied: a gradual decline at 50°C/min resulted in the formation of LK-99 sample 1 (denoted as S1), while a faster cooling rate of 10°C/min produced LK-99 sample 2 (denoted as S2). The final procedure for the LK-99 with different cooling rates was depicted in Figure 1(a). As mentioned in the introduction, the LK-99 sample might contain considerable impurities. Moreover, during its synthesis, there is a possibility that $Cu_2S$ could form and be present in the LK-99, potentially influencing the electrical and magnetic measurements [14,16,17]. To validate the properties of LK-99 with and without the $Cu_2S$ impurity, samples were immersed in an ammonia solution NH$_3$·H$_2$O for 3 hours. The Cu$_2$S will undergo dissolution in the solution. The samples that have been immersed in the ammonia solution are indicated with an asterisk in the article, as S1* or S2*. Figure 1(b) displays images of S1 before immersion in the ammonia solution. The image of S1 during its immersion in the ammonia solution is shown in Figure 1(c). It can be seen that the ammonia solution appears in light blue due to the presence of copper ions. The phase purity and the crystal structure of obtained LK-99 samples were characterized by powder x-ray diffraction (Bruker D2 phaser) measurements with Cu-$K_\alpha$ radiation on samples.

**Transport measurements.** For electrical transport studies, we resized the LK-99 samples to dimensions of roughly 1.5 × 1.0 × 0.3 mm$^3$. We then affixed four leads on the samples using silver paint. The resistivity ρ was measured via the conventional DC four-probe method. Additionally, the magnetization of the LK-99 samples was measured by using a superconducting quantum interference device system (MPMS from Quantum Design).

Results

**X-ray diffraction characterizations of LK-99 grown under different conditions.** The crystalline structures of the LK-99 samples with and without Cu$_2$S were analyzed via the XRD measurement. The x-ray $\theta$–$2\theta$ diffraction spectra shown in Figure 2(a) reveal characteristic peaks of the LK-99 sample, S1, which was not immersed in an ammonia solution. These peaks align with findings previously reported by [1,2,18]. On the other hand, Figure 2(b) showcases the LK-99 sample immersed in the ammonia solution, S1*. It displays similar characteristic XRD peaks, with an absent peak around 37° when compared to the XRD result of S1. The notable peak observed at 37°could suggest the (1,0,2) phase of Cu$_2$S [18-20] in the S1.

**Magnetic properties of LK-99.** The temperature-dependent magnetization of the samples was measured in zero-field-cooled (ZFC) and field-cooled (FC) modes with an applied field of $H$ = 200 Oe as shown in Figure 3. As illustrated in the figure, all samples exhibit a low magnetization $M$, approximately $10^{-5}$ emu/g, and maintain a negative magnetization over the temperature range from 400 K to 20 K in both ZFC and FC. The ZFC magnetization of S1 demonstrates a decreasing trend within the range from 400 to 360 K. As the temperature continues to decrease, the magnetization begins to rise gradually. Upon approaching 20 K, the magnetization increases sharply and becomes positive. The FC result of S1 is similar to ZFC but shows a separation of ZFC and FC curves, which may arise from the pinning effect of some impurities. S1* has a slightly stronger negative magnetization. The magnetization's response to temperature was similar to that of S1, however, with the ZFC and FC of S1* being nearly the same. Despite the negative magnetization observed in LK-99 samples, the intrinsic low magnetic response and the paramagnetic behavior observed near 2 K indicated that our LK-99 might not be a superconductor.

**Basic resistive properties of LK-99.** The temperature-dependent resistivity of our samples is shown in Figure 4. The range of the measurement spans from 90 K to 420 K. In Figure 4(a), the resistivity of S1 shows a decline as the temperature decreases from 400 K. At 380 K, a sharp transition is observed as resistivity increases from 0.1 to 0.4 Ω cm. Subsequently, the resistivity demonstrates semiconductor behavior, increasing as the temperature decreases. The magnitude of the resistivity of S1* is approximately the same as S1. Besides, the transition previously observed at 384 K in S1 cannot be observed in S1*. The temperature-dependent resistivity of S2 and S2*, presented in Figure 4(b), exhibits similar characteristics to those of S1 and S1*. The apparent transition near 384 K in S2 is conspicuously absent in S2*, suggesting a consistent conclusion. The result shows that the sharp transition of the resistivity near 380 K could be attributed to the $Cu_2S$ in LK-99. The resistivity transition observed above room temperature by the Korean team, which they claimed as evidence of a room-temperature superconductor, might also be caused by $Cu_2S$, thus leading to an erroneous conclusion.

**Discussion**

Since the so-called superconducting-like behavior in LK-99 is most likely due to a reduction in resistivity caused by the structural transition of $Cu_2S$ at around 385 K, it is very worthwhile to study the LK-99 samples with the removal of $Cu_2S$. In our LK-99 samples, XRD analysis revealed a $Cu_2S$ phase, as indicated by a distinct peak at $2\theta$ of approximately 37°. This phase was removed with an ammonia solution treatment as evidenced by the existence of the blue $Cu^{+2}$ ions in the solution and the disappearance of $Cu_2S$ peak in XRD patterns. Magnetic measurements for $Cu_2S$-removed LK-99 samples across 10 K to 400 K revealed a consistent negative trend. The observed magnetic magnetization deviates from the expected superconductor, showing a diamagnetic nature at room temperature. Additionally, the resistivity transition above room temperature in LK-99 could be attributed to the influence of $Cu_2S$. The Korean research team might have been misinterpreted as indicating of a room-temperature superconductor. Details of electrical and magnetic properties for a clean $Pb_9Cu(PO_4)_6O$ phase require further studies.

In summary, we measured transport and magnetic properties of LK-99 samples with and without the removal of $Cu_2S$. It is found that the $Cu_2S$ phase can be effectively removed by ammonia solution. The superconducting-like behavior in LK-99 should originate from a magnitude reduction in resistivity caused by the presence of $Cu_2S$. Thus, it might not be possible to consider LK-99 as a room-temperature superconductor but to be a diamagnetic semiconductor, and the quest for a room-temperature superconductor remains a challenge in the research field.

**Data availability**

The data supporting the findings of this study are available on reasonable request.

**Acknowledgments**

The authors thank the National Science and Technology Council (NSTC) of the Republic of China for financial support under grant number NSTC 112-2112-M-002-028.

**Authors contributions**

Z.L. synthesized the polycrystalline LK-99 samples and performed XRD & electrical transport measurements. C.W.L and I.N.C. performed the magnetization measurements. C.T.C assisted in the electrical transport measurements. Z.L. and L.M.W. analyzed the data. C.W.L and L.M.W. wrote the manuscript.

**Competing interests**

The authors declare no competing interests.

**Additional information**

**Correspondence** and requests for materials should be addressed to L.M.W.

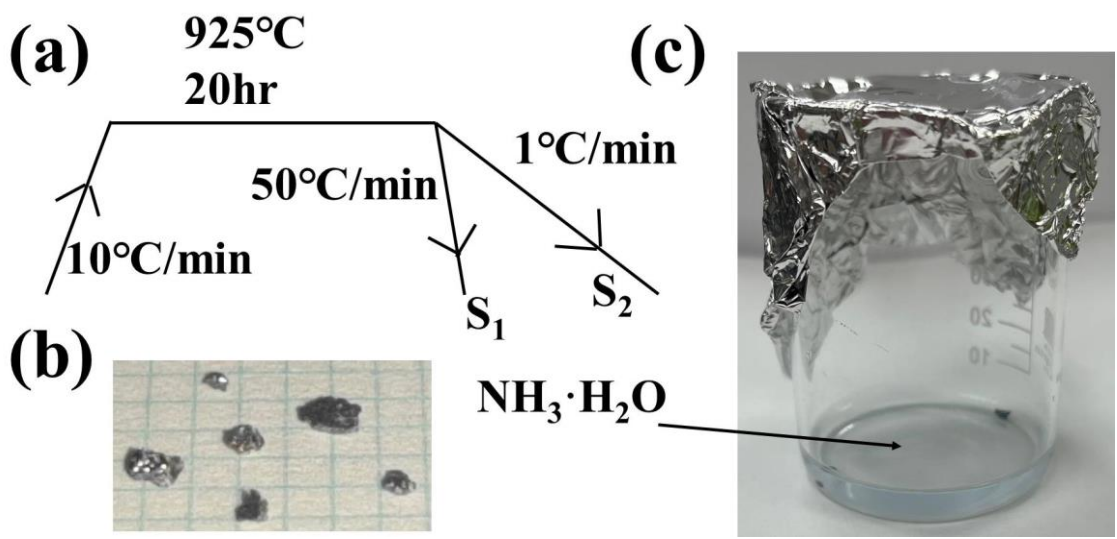

**Fig. 1 Synthesis process of LK-99. a** Schematic process of final procedure with different cooling rates denoted as S1 and S2. **b** The images of S1. **c** The image of the S1 immersed in ammonia solution, showing a light blue color of $Cu^{2+}$.

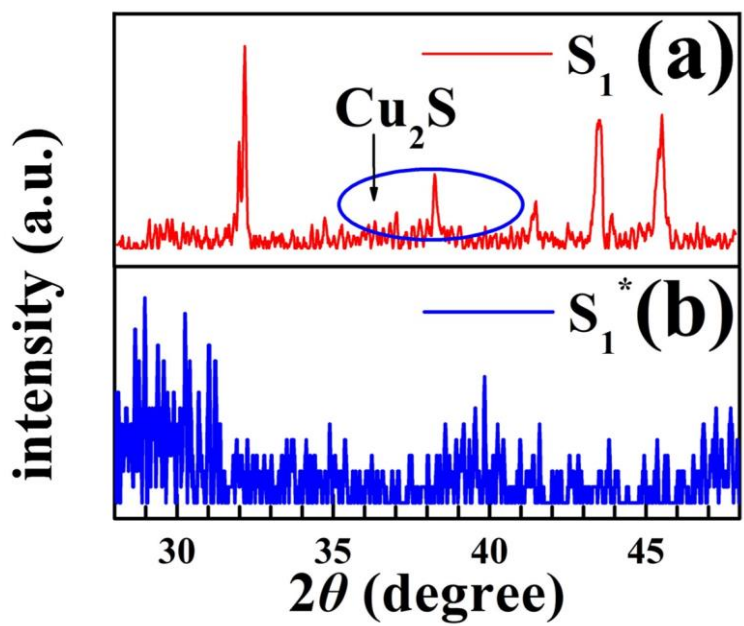

**Fig. 2 X-ray diffraction characterizations of LK-99. a** X-ray $\theta$–$2\theta$ diffraction spectrum for the S1. **b** X-ray $\theta$–$2\theta$ diffraction spectrum for the S1*.

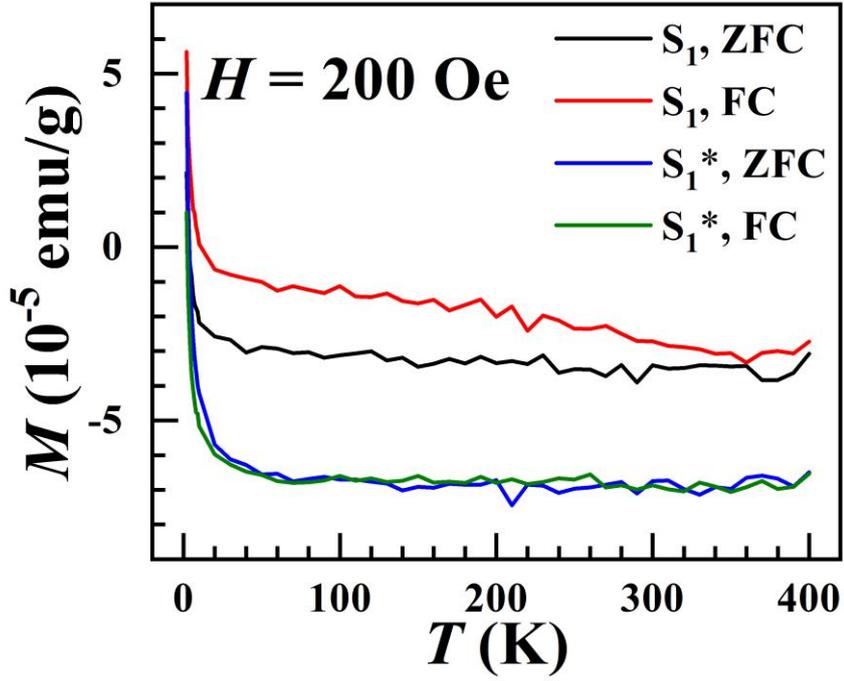

**Fig. 3 Magnetic properties of LK-99.** ZFC and FC magnetizations in $H$ = 200 Oe for samples S1, S1*, S2 and S2*.

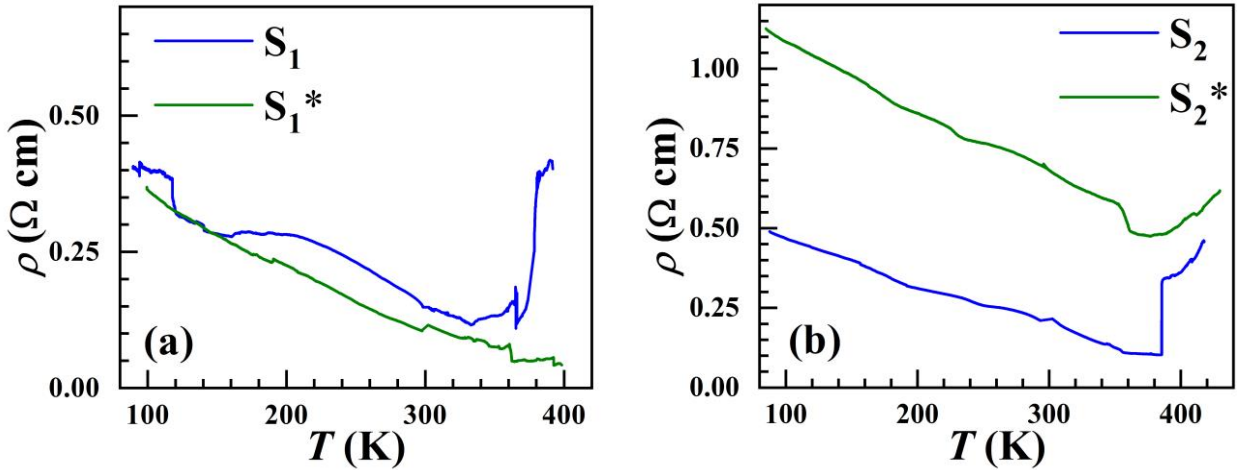

**Fig. 4 Basic resistive properties of LK-99. a** Resistivity of S1 shown in blue, and resistivity of S1* depicted in green. **b** Resistivity of S2 shown in blue, and resistivity of S2* depicted in green. The apparent transition near 380 K is conspicuously absent in S1* and S2*.